\documentclass[aps,prl,groupaddress,twocolumn,floatfix,a4paper]{revtex4}

\usepackage{graphicx,graphics,epsfig}   
\usepackage{dcolumn}    
\usepackage{bm}         
\usepackage{amsmath}    
\usepackage{verbatim}   
\usepackage{color}      
\usepackage{subfigure}  
\usepackage{times,natbib}
\usepackage{amsmath,amsfonts,amssymb,graphics,graphicx,epsfig,color,times,natbib}

\newcommand{\PPR}[0]{P^{\text{PR}}}
\newcommand{\PPRanti}{P^{\overline{\text{PR}}}}

\newcommand{\Pc}[0]{P^{\text{c}}}
\newcommand{\Pa}[0]{P^{\text{a}}}

\newcommand{\be}{\begin{equation}}
\newcommand{\ee}{\end{equation}}
\newcommand{\ba}{\begin{eqnarray}}
\newcommand{\ea}{\end{eqnarray}}
\newcommand{\ban}{\begin{eqnarray*}}
\newcommand{\ean}{\end{eqnarray*}}

\begin{document}


\title{Non-locality distillation and post-quantum theories with trivial communication complexity}

\author{Nicolas Brunner}%
\email{n.brunner@bristol.ac.uk}
\affiliation{ H.H. Wills Physics Laboratory, University of Bristol, Tyndall Avenue, Bristol, BS8 1TL, United
Kingdom }
\author{Paul Skrzypczyk}
\affiliation{ H.H. Wills Physics Laboratory, University of Bristol, Tyndall Avenue, Bristol, BS8 1TL, United
Kingdom }

\date{\today}

\begin{abstract}
We first present a protocol for deterministically distilling non-locality, building upon a recent result of Forster et al. [Phys. Rev. Lett. 102, 120401 (2009)]. Our protocol, which is optimal for two-copy distillation, works efficiently for a specific class of post-quantum non-local boxes, which we term correlated non-local boxes. In the asymptotic limit, all correlated non-local boxes are distilled to the maximally non-local box of Popescu and Rohrlich. Then, taking advantage of a result of Brassard \textit{et al.} [Phys. Rev. Lett. 96, 250401 (2006)] we show that all correlated non-local boxes make communication complexity trivial, and therefore appear very unlikely to exist in nature. Astonishingly, some of these non-local boxes are arbitrarily close to the set of classical correlations. This result therefore gives new insight to the problem of why quantum non-locality is limited.
\end{abstract}

\maketitle

By performing measurements on an entangled quantum state, two distant observers can establish correlations that are non-local, in the sense of violating a Bell inequality \cite{bell64}. The non-locality of these correlations is what makes them so powerful for processing information. Nevertheless quantum non-locality is bounded, as found by Tsirelson \cite{tsirelson2}. In a seminal paper, Popescu and Rohrlich \cite{PR} showed that, surprisingly, this bound is not a consequence of relativity, i.e. there exist correlations respecting the no-signaling principle that are more non-local than those of quantum mechanics. Identifying what physical principle limits quantum non-locality is an important problem in the foundations of quantum mechanics.

Recently, it was suggested that Tsirelson's bound could be a consequence of the information theoretic properties of general non-signalling theories \cite{john,brassardNP,sanduNP}, and several works underwent the task of demonstrating a separation between quantum mechanics and post-quantum theories. In particular, van Dam \cite{vanDam} showed that the availability of Popescu-Rohrlich boxes \cite{PR,barrett}, the paradigmatic non-local box, makes communication complexity trivial, while quantum mechanics does not \cite{IP}. More generally, it is strongly believed that theories in which communication complexity is trivial are very unlikely to exist. The result of \cite{vanDam} was subsequently extended by Brassard \textit{et al.} \cite{brassard}, to a set of post-quantum models known as isotropic non-local boxes. However a gap subsisted, in the sense that their proof applied only to isotropic boxes violating the Clauser-Horne-Shimony-Holt (CHSH) \cite{chsh} inequality by more than $\mathcal{B}_{cc} \equiv 4\sqrt{2/3} \approx 3.266$; whereas quantum correlations are limited by Tsirelson's bound $\mathcal{B}_{Q} \equiv 2\sqrt{2}\approx 2.828$. Moreover, Linden \textit{et al.} \cite{noah} showed that all post-quantum isotropic boxes allow for non-local computation, thus indicating the first tight separation with quantum correlations. Finally, Tsirelson's bound was also found to appear (in a very unexpected way) in the study of the dynamical process of non-locality swapping \cite{emergence,couplers}, and in theories with relaxed uncertainty relations \cite{verSteeg}.

From a more general perspective, the problem of finding why quantum non-locality is bounded does not reduce only to recovering Tsirelson's bound, but more generally to recovering the full boundary of the quantum set of correlations \cite{TLM,miguel}. Indeed it is not so that all correlations below Tsirelson's bound are attainable in quantum mechanics (see Fig. 1); in fact there exist post-quantum correlations that are arbitrarily close to the set of classical correlations. Again, it is a very natural question to ask why such correlations would be unlikely to exist in nature.

\begin{figure}[b]
\includegraphics[width=0.75\columnwidth]{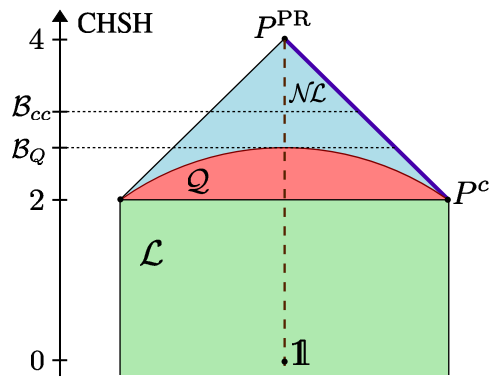} \caption{The set of quantum non-local correlations $\mathcal{Q}$ is a strict subset of the correlations constrained solely by the non-signaling principle $\mathcal{NL}$. In particular, there are post-quantum correlations above, but also below Tsirelson's bound $\mathcal{B}_Q=2\sqrt{2}$. Post-quantum correlations violating the CHSH inequality by more than $\mathcal{B}_{cc}\approx3.266$, are very unlikely to exist, since they make communication complexity trivial. Here we will show that a similar conclusion holds for correlated non-local boxes (bold line). The dashed vertical line represents isotropic non-local boxes.}\label{poly1}
\end{figure}

In the present paper, we partially answer this question by showing that a particular class of post-quantum correlations, which we term ``correlated non-local boxes", make communication complexity trivial. Astonishingly, some of these correlated non-local boxes give an arbitrarily small violation of the CHSH inequality, and are arbitrarily close to the set of classical and quantum correlations. To demonstrate this result we construct a protocol for deterministically distilling non-locality.

We build upon a recent breakthrough of Forster et al. \cite{foster}, who presented the first non-locality distillation protocol. The key element in their study was to look beyond isotropic boxes \cite{dejan,toni_pur}. However, since their protocol can distill only up to $\text{CHSH}=3$, it does not allow one to reach the bound $\mathcal{B}_{cc}$. Here we present a protocol that distills any correlated non-local box (in the asymptotic limit) to the maximally non-local box, i.e. to the PR box ($\text{CHSH}=4$). Thus our protocol, which happens to be optimal for two-copy distillation, can distill above the bound $\mathcal{B}_{cc}$ and therefore implies that all correlated non-local boxes collapse communication complexity.

\emph{Preliminaries.} Our study focuses on the case of the CHSH scenario, which we will describe here in terms of non-local boxes. Two distant parties, Alice and Bob, share a non-local box. Each party is allowed to input one bit into the box and gets one output bit: Alice inputs $x \in \{0,1\}$ and gets outcome $a\in \{0,1\}$; Bob inputs $y \in \{0,1\}$ and gets outcome $b\in \{0,1\}$. Every non-local box is then characterized by a set of 16 joint probabilities $P(ab|xy)$. Here we will focus on the situation where the output bits $a,b$ are unbiased.

The PR box is characterized by the following probability distribution:
\begin{equation}\label{PR}
\PPR (ab|xy) =
\begin{cases}
\frac{1}{2} & \text{$ a \oplus b = xy$} \\
0 & \text{otherwise}
\end{cases}
\end{equation} where $\oplus$ is addition modulo 2. Moreover we will also consider the ``correlated local state", for which the outputs are always perfectly correlated (independent of the inputs), i.e.

\begin{equation}\label{PR}
\Pc (ab|xy) =
\begin{cases}
\frac{1}{2} & \text{$ a \oplus b = 0$} \\
0 & \text{otherwise}
\end{cases}
\end{equation}

Next, we will mainly focus on the family of "correlated non-local boxes", defined as follows:

\ba\label{family} \PPR_{\epsilon} = \epsilon \PPR + (1-\epsilon) \Pc \ea where $0 \leq \epsilon \leq 1$. The box $\PPR_{\epsilon}$ has CHSH value of $\text{CHSH}_{i} = 2(\epsilon + 1)$. Here the CHSH polynomial is given by $E_{00}+E_{01}+E_{10}-E_{11}$, where $E_{xy}=P(a=b|xy)-P(a \neq b|xy)$ is the correlator for the pair of measurements $x,y$.

\emph{Distillation protocol.} Now we present a non-locality distillation protocol that works deterministically for correlated non-local boxes \eqref{family}. More precisely our protocol takes two copies of any box $\PPR_{\epsilon}$ with $0<\epsilon<1$ to a correlated non-local box $\PPR_{\epsilon'}$ with $\epsilon' > \epsilon$, thus distilling non-locality. Moreover in the asymptotic regime of many copies, any box $\PPR_{\epsilon}$ with $\epsilon>0$ is distilled arbitrarily close to the PR box. This protocol is optimal for deterministic two copy distillation.

The protocol works as follows (see Fig. 2). Alice and Bob share two boxes. Let us denote $x_i$ the value that Alice inputs into box $i$, and similarly $y_j$ the value Bob inputs into box $j$. The output bits of box number $k$ are then denoted $a_k,b_k$. Alice proceeds as follows: $x_1=x$, $x_2=xa_1$, and she outputs finally $a= a_1\oplus a_2$. Bob proceeds in a similar way: $y_1=y$, $y_2=yb_1$, and he outputs finally $b= b_1\oplus b_2$.

\begin{figure}[t]
\includegraphics[width=0.58\columnwidth]{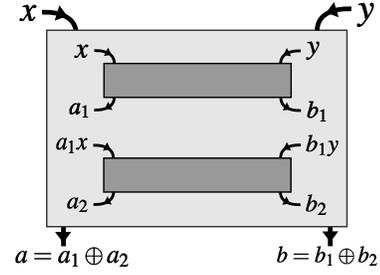} \caption{Protocol for two-copy deterministic non-locality distillation.}\label{Protocol}
\end{figure}

Now we prove that the protocol distills two copies of any box $\PPR_{\epsilon}$ with $0<\epsilon<1$. The initial two box state is
\ba\nonumber \PPR_\epsilon \PPR_\epsilon &=& \epsilon^2 \PPR \PPR + \epsilon(1-\epsilon) (\PPR \Pc + \Pc \PPR) \\ & & + (1-\epsilon)^2 \Pc \Pc \ea Let us now compute the final box, after the above distillation protocol has been applied. It is convenient to proceed step by step. The notation $P_iP'_i \rightarrow P_f$ means that the protocol takes two initial boxes, $P_i$ and $P'_i$, to one copy of the final box $P_f$.

\begin{itemize}
  \item \emph{$\PPR \PPR \rightarrow \PPR$ .} For the first box we have $a_1 \oplus b_1 = xy$, implying $b_1= a_1 \oplus xy$. For the second box, we have $a_2\oplus b_2 = xa_1y b_1 = xy a_1 (a_1 \oplus xy)=0$. So the outputs satisfy the relation $a \oplus b = a_1 \oplus a_2 \oplus b_1 \oplus b_2 = xy$.
  \item \emph{$\PPR \Pc \rightarrow \PPR$ .} For the first box we have $a_1 \oplus b_1 = xy$. For the second box we have $a_2\oplus b_2=0$ independently of the inputs. So finally the outputs satisfy $a \oplus b = xy$.
  \item \emph{$\Pc \PPR \rightarrow \frac{1}{2}(\PPR+\Pc)$ .} For the first box we have $a_1 \oplus b_1 = 0$, implying $b_1= a_1$. For the second box, we have $a_2 \oplus b_2 = xya_1b_1= xy a_1$, where $a_1$ is random. When $a_1=0$, one gets $a\oplus b=0$; when $a_1=1$, one gets $a\oplus b=xy$.
  \item \emph{$\Pc \Pc \rightarrow \Pc$ .} Here we have that $a_1 \oplus b_1 = 0$ and $a_2 \oplus b_2 = 0$. Therefore $a \oplus b = 0$.
\end{itemize}

\begin{figure}[b]
\includegraphics[width=0.95\columnwidth]{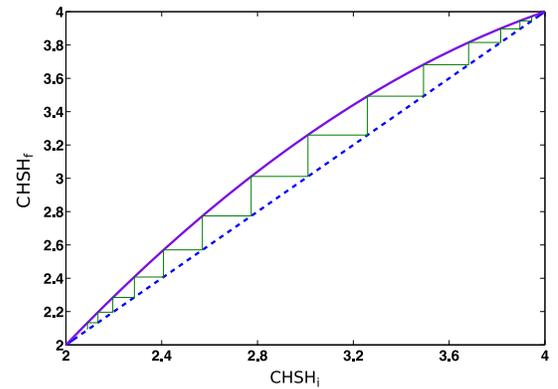} \caption{Non-locality distillation of correlated non-local boxes with our protocol. The graph shows the $\text{CHSH}_f$ value of the final box as a function of its initial $\text{CHSH}_i$ value. The dashed straight line, $\text{CHSH}_f=\text{CHSH}_i$, is given as a reference. In the asymptotic regime, any non-local box $\PPR_{\epsilon}$ is distilled arbitrarily close to the PR box (CHSH=4). The thin line (steps) shows the distillation of an initial box under successive iterations of the protocol.}\label{Graph}
\end{figure}

So the final box, after distillation, is given by

\ba \PPR_{\epsilon'} = \frac{\epsilon}{2}(3 - \epsilon) \PPR +  [1-\frac{\epsilon}{2}(3 - \epsilon)] \Pc \ea Its CHSH value is $\text{CHSH}_{f} = 3\epsilon-\epsilon^2+2$. Imposing that $\text{CHSH}_f > \text{CHSH}_i$ implies that $\epsilon - \epsilon^2 >0$, a condition always satisfied for $0<\epsilon<1$. Therefore the protocol distills any box in the family \eqref{family}, as shown in Fig. 3.

Note that it is convenient to view this distillation protocol as a discrete map. Starting with $2^n$ copies of a box $\PPR_{\epsilon}$, the final box $\PPR_{\epsilon_n}$ is characterized by the parameter $\epsilon_n$ which is the $n^{th}$ iteration of the map

\ba T(\epsilon) = \epsilon'= \frac{\epsilon}{2}(3 - \epsilon) \ea Notably the fixed points of the map (i.e. the asymptotic behaviour) are $\epsilon=0$, i.e. the local box $\Pc$, and $\epsilon=1$, i.e. the $\PPR$ box. The stability of both of these points can be checked by finding the eigenvalues of the Jacobian at the fixed points. Here the problem being one-dimensional, this reduces to calculating $\lambda = \frac{d T}{d \epsilon}$. For the first fixed point, $\epsilon=0$ ($\Pc$), we find $\lambda=\frac{3}{2}$, indicating that the point is repulsive. For the second fixed point, $\epsilon=1$ ($\PPR $), we find $\lambda=\frac{1}{2}$, indicating that the point is attractive.

In general, a distillation protocol can be viewed as a way of (classically) wiring the boxes together. In ref \cite{short}, a classification of all possible wirings has been given. It was shown that (consistent) wirings form a convex set, a polytope. There are 82 extremal wirings, which can be classified in five classes: deterministic, one-sided, XOR, AND, and sequential. A distillation protocol is then a choice of four wirings; one for each input of Alice and Bob. Our protocol mixes sequential and XOR wirings. We checked that it is optimal for deterministic two copy distillation (with classical wirings). Note that the protocol of \cite{foster} is based solely upon XOR wirings, of which all are suboptimal for two-copy distillation. These protocols can distil non-locality only up to $\text{CHSH}=3$, and therefore do not allow one to reach the bound $\mathcal{B}_{cc}$.

\emph{Distillation in a section of the polytope.} In this section, we study how our distillation protocol works for more general non-local boxes, of the form

\ba\label{noisy} \PPR_{\xi,\gamma} \equiv \xi \PPR + \gamma \Pc + (1- \xi - \gamma )  \PPRanti \ea
with $\xi,\gamma \geq 0$ and $\xi + \gamma \leq 1$, and where $\PPRanti$ denotes the anti-PR box, given by the relation $a \oplus b = xy \oplus 1$. Note that the family of boxes $\PPR_{\xi,0}$ are the isotropic PR boxes; the family of boxes $\PPR_{\xi,1-\xi}$ are the correlated non-local boxes \eqref{family}.

In order to compute the final box after distillation of two initial boxes $\PPR_{\xi,\gamma}$, the following relations are needed: $\PPR \PPRanti \rightarrow \PPRanti$, $\PPRanti \PPR  \rightarrow \frac{1}{2}(\PPRanti + \Pa)$, $\PPRanti \Pc  \rightarrow \PPRanti$, $\PPRanti \PPRanti \rightarrow \frac{1}{2}(\PPR+\Pc)$, $\Pc \PPRanti   \rightarrow \frac{1}{2}(\PPRanti + \Pa)$, where $\Pa$ is the anti-correlated (local) box given by $a\oplus b=1$ for all $x,y$. Here it is convenient to use the non-convex decomposition $\Pa =  \PPR + \PPRanti - \Pc$, since $\Pa$ lies in the plane defined by $\PPR$, $\PPRanti$ and $\Pc$, but is not contained in their convex hull. Note that from now on we must have $-1 \leq \gamma\leq 1$ and $0 \leq \xi+\gamma \leq 1$.

The distillation protocol then implements the map

\ba T_1(\xi,\gamma)  &=& \xi'= \xi^2 +\frac{3}{2}\xi\gamma + \frac{1}{2}(1-\xi-\gamma) \\\nonumber T_2(\xi,\gamma)  &=& \gamma' = \xi^2 + 2\gamma^2 + \frac{5}{2}\xi\gamma - \frac{3}{2}(\xi+\gamma)+\frac{1}{2} \ea and the final box is given by $\PPR_{\xi',\gamma'}$. This map has three fixed points: $\PPR_{1,0}= \PPR$, $\PPR_{0,1}= \Pc$, and $\PPR_{1/2,0}= \openone$, where $\openone$ is the fully mixed box given by $a\oplus b = c$ and $c$ is a random bit. Again, the stability of these points can be studied by computing the eigenvalues of the Jacobian at each fixed point. We find that $\openone$ is an attractor, $\Pc$ is an unstable point, and $\PPR$ is an unstable saddle point. For the correlated non-local boxes \eqref{family}, $\PPR$ is indeed an attractive point, but in all other directions it becomes repulsive. This implies that the only boxes which are distilled in the asymptotic regime are correlated non-local boxes. In Fig. 4, we highlight the set of boxes that are distilled in a single iteration of the protocol; note that these do not include the isotropic boxes, since they cannot be deterministically distilled by a two copy protocol \cite{toni_pur}.

\emph{Trivial communication complexity.} It was shown in Ref \cite{brassard} that isotropic non-local boxes violating the CHSH inequality by more than $\mathcal{B}_{cc}  \approx 3.266$ make communication complexity trivial. In fact this result also holds for non-isotropic boxes violating the CHSH inequality by more than $\mathcal{B}_{cc}$, since there exists a depolarizing protocol, found by Masanes et al. \cite{NS} and independently by Short \cite{toni_pur}. This protocol, which can be performed locally and without communication, maps any non-isotropic box to an isotropic one, leaving its CHSH value unchanged. The protocol works as follows. Alice and Bob generate three random bits $\alpha,\beta,\gamma$, and modify their inputs and outputs as follows: $x\rightarrow x \oplus \alpha$ , $y \rightarrow y \oplus \beta$ , $a \rightarrow a\oplus \beta x \oplus \alpha\beta \oplus \gamma$ , $b\rightarrow b \oplus \alpha y \oplus  \gamma$.

\begin{figure}[t]
\includegraphics[width=\columnwidth]{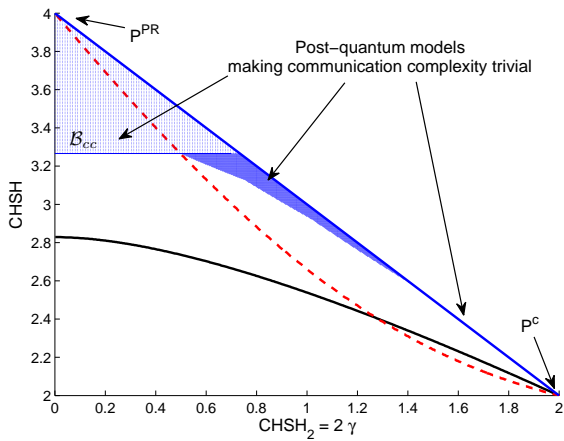} \caption{(Color online) Section of the non-signaling polytope given by non-local boxes $\PPR_{\xi,\gamma}$. The blue line and the shaded (blue) areas represent the set of post-quantum boxes that collapse communication complexity. The area shaded in light blue was identified in \cite{brassard}, while the area in dark blue has been identified in the present paper. Astonishingly, some of the correlated non-local boxes are arbitrarily close to the local state $\Pc$, yet they still collapse communication complexity. Non-local boxes above the dashed (red) curve are distilled in one iteration of the protocol. The solid curve is the quantum boundary \cite{TLM}.}\label{Poly}
\end{figure}

Now it is clear that any correlated non-local box $\PPR_\epsilon$ \eqref{family} violating the CHSH inequality by less than $\mathcal{B}_{cc}$ can be distilled using our protocol to a box giving a violation above this bound. Therefore, any box $\PPR_\epsilon$ with $\epsilon>0$ makes communication complexity trivial. This result is indeed astonishing since a box $\PPR_{\epsilon}$ can be arbitrarily close to the local set (and consequently also to the quantum set). This shows again that the CHSH violation is an inadequate measure of non-locality \cite{EntVsNL,foster}.

More generally, all non-local boxes $\PPR_{\xi,\gamma}$ that can be distilled (in any number of iteration of the protocol) above $\mathcal{B}_{cc}$ make communication complexity trivial. These boxes form a region highlighted in Fig. 4. Importantly this region is not singular; it has always a finite (non-zero) width, so that it is robust against small perturbations. Indeed, the amount of tolerable perturbation depends on the CHSH value, and the system becomes more sensitive to perturbation as we get closer to the CHSH limit of 2, but even there, for small enough, yet finite, perturbations, the system is robust.

\emph{Conclusion.} We started by presenting a protocol for distilling non-locality, that works for the family of correlated non-local boxes. All of these boxes, even those giving arbitrarily small violation of the CHSH inequality, can be distilled to the PR box in the asymptotic limit. Furthermore, the existence of this protocol has important implications from the point of view of communication complexity in post-quantum theories: all non-local boxes that can be distilled above CHSH=$\mathcal{B}_{cc}\approx 3.266$ using our protocol make communication complexity trivial. Astonishingly, some of these non-local boxes are arbitrarily close to the set of local (and quantum) boxes, yet they still collapse communication complexity. In this sense, these boxes appear to be as non-local as the PR box, and therefore seem very unlikely to exist in nature. This result provides a partial answer to the question of why quantum non-locality is also bounded below Tsirelson's bound, in regions of the polytope close to the local set of correlations.

In the future it would be interesting to find better distillation protocols, which might witness a larger set of post-quantum non-local boxes that make communication complexity trivial. Of particular interest would be protocols based on generalized joint measurements, so-called couplers \cite{emergence,couplers}, which may enhance distillation, in analogy to the advantage given by joint measurements in quantum mechanics. Indeed the ultimate goal of this line of research would be to obtain a tight separation between quantum and any post-quantum correlations, eventually leading to a new information-theoretic axiom for quantum mechanics \cite{brassardNP}.


\emph{Acknowledgements.} The authors are grateful to J.~E. Allcock, A.~R.~U. Devi, W. Matthews, and S. Popescu for insightful discussions. P. S. acknowledge support through the UK EPSRC project `QIP IRC'. N. B. acknowledges financial support by the Swiss National Science Foundation (SNSF).

\bibliographystyle{prsty}
\bibliography{C:/BIB/thesis}

\end{document}